\documentclass[pdflatex,sn-mathphys-num]{sn-jnl}

\usepackage{graphicx}%
\usepackage{multirow}%
\usepackage{amsmath,amssymb,amsfonts}%
\usepackage{amsthm}%
\usepackage{mathrsfs}%
\usepackage[title]{appendix}%
\usepackage{xcolor}%
\usepackage{textcomp}%
\usepackage{manyfoot}%
\usepackage{booktabs}%
\usepackage{gensymb}
\usepackage{algorithm}%
\usepackage{algorithmicx}%
\usepackage{algpseudocode}%
\usepackage{listings}%
\usepackage{booktabs}
\usepackage{multirow}
\usepackage{xr-hyper}

\makeatletter
\newcommand*{\addFileDependency}[1]{ 
  \typeout{(#1)}
  \@addtofilelist{#1}
  \IfFileExists{#1}{}{\typeout{No file #1.}}
}
\makeatother

\newcommand*{\myexternaldocument}[3][supp-]{%
  \externaldocument[#1]{#2/#3}%
    \addFileDependency{#3.tex}%
    \addFileDependency{#2/#3.aux}%
}
\myexternaldocument[supp-]{build}{sn-supplement}

\theoremstyle{thmstyleone}%
\theoremstyle{thmstyletwo}%

\theoremstyle{thmstylethree}%

\begin{document}

\title[Article Title]{Hybrid Physics-Machine Learning Models for Quantitative Electron Diffraction Refinements}


\author[1]{\fnm{Shreshth A.} \sur{Malik}}
\equalcont{These authors contributed equally to this work.}

\author*[1,2]{\fnm{Tiarnan A.S.} \sur{Doherty}}\email{td404@cam.ac.uk}
\equalcont{These authors contributed equally to this work.}

\author[2]{\fnm{Benjamin} \sur{Colmey}}

\author[3]{\fnm{Stephen J.} \sur{Roberts}}

\author*[1]{\fnm{Yarin} \sur{Gal}}\email{yarin.gal@cs.ox.ac.uk}

\author*[2]{\fnm{Paul A.} \sur{Midgley}}\email{pam33@cam.ac.uk}

\affil[1]{\orgdiv{OATML, Department of Computer Science}, \orgname{University of Oxford}, \orgaddress{\street{Wolfson Building, Parks Rd}, \city{Oxford}, \postcode{OX1 3QG},  \country{United Kingdom}}}

\affil[2]{\orgdiv{Department of Materials Science and Metallurgy}, \orgname{University of Cambridge}, \orgaddress{\street{27 Charles Babbage Rd}, \city{Cambridge}, \postcode{CB3 0FS}, \country{United Kingdom}}}

\affil[3]{\orgdiv{MLRG, Department of Engineering Science}, \orgname{University of Oxford}, \orgaddress{\street{Eagle House, Walton Well Road}, \city{Oxford}, \postcode{OX2 6ED},  \country{United Kingdom}}}

\abstract{
 
High-fidelity electron microscopy simulations required for quantitative crystal structure refinements face a fundamental challenge: while physical interactions are well-described theoretically, real-world experimental effects are challenging to model analytically.
To address this gap, we present a novel hybrid physics–machine learning framework that integrates differentiable physical simulations with neural networks. By leveraging automatic differentiation throughout the simulation pipeline, our method enables gradient-based joint optimization of physical parameters and neural network components representing experimental variables, offering superior scalability compared to traditional second-order methods. We demonstrate this framework through application to three-dimensional electron diffraction (3D-ED) structure refinement, where our approach learns complex thickness distributions directly from diffraction data rather than relying on simplified geometric models. This method achieves state-of-the-art refinement performance across synthetic and experimental datasets, recovering atomic positions, thermal displacements, and thickness profiles with high fidelity. The modular architecture proposed can naturally be extended to accommodate additional physical phenomena and extended to other electron microscopy techniques. This establishes differentiable hybrid modeling as a powerful new paradigm for quantitative electron microscopy, where experimental complexities have historically limited analysis.}




\maketitle

\section{Introduction}\label{sec:intro}
Recent advances in scientific machine learning have demonstrated remarkable potential to complement traditional physics-based simulations across diverse domains of computational science. Hybrid physics–machine learning approaches, which combine interpretable physical models with the expressive power of neural networks~\citep{goodfellow2016deep, KrizhevskySH12, he2016deep, silver2017mastering, schrittwieser2020mastering}, offer a promising avenue for addressing complex phenomena that remain challenging for purely analytical methods. 
By leveraging neural networks as universal function approximators~\citep{hornik1989multilayer} to model intricate relationships while preserving the interpretability and constraints of established physical frameworks, hybrid approaches bridge the gap between first-principles theory and experimental reality.

These hybrid methodologies have shown substantial promise in fields ranging from dynamical systems~\citep{de2018end, allen2022physical, pfaff2020learning} and atmospheric physics~\citep{arcomano2022hybrid} to quantum physics~\citep{deringer2019machine, schoenholz2020jax}, where learned components augment traditional simulators to improve realism, accuracy, or computational efficiency. A key enabler of these innovations is differentiable physics; the reformulation of physical simulations to be compatible with automatic differentiation frameworks \citep{baydin2018automatic, pennington_third-dimension_2014, van_den_broek_method_2012, diederichs_exact_2024}. By making every step of a physical simulation differentiable, gradients of physical quantities with respect to model parameters can be computed efficiently, enabling the use of gradient-based optimization methods to simultaneously refine both physical parameters and neural network components. 

Electron microscopy represents an ideal setting for the use of a hybrid modeling approach because while the elastic interaction of the electron beam with matter is well defined by established physical theory ~\citep{humphreys_scattering_1979, kirkland_advanced_2010}, real experiments involve sample morphologies, inelastic scattering, electron beam damage, atomic scale defects and other sample heterogeneity that are notoriously difficult to parameterize. Omitting these contributions in simulations imposes a systematic ceiling on the precision and accuracy with which structural and compositional information can be recovered from experimental data. However, to the best of our knowledge, there has not yet been a demonstration of the utility of hybrid modeling approaches when applied to any electron microscopy technique.

\begin{figure}
    \centering
    \includegraphics[width=1.0\linewidth]{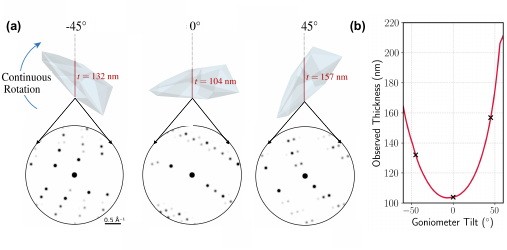}
    \caption{Schematic of the 3D-ED method, in which an irregularly shaped crystal is continuously rotated under the electron beam while diffraction patterns are recorded. (a) Three representative orientations of a quartz crystal during rotation, each shown with corresponding simulated dynamical diffraction patterns. (b) Crystal thickness along the beam direction as a function of tilt angle, with selected orientations marked.}
    \label{fig:crystaltilt}
\end{figure}

One such electron microscopy modality that could benefit substantially from a hybrid modeling approach is the field of three dimensional electron diffraction (3D-ED), where a crystal is rotated relative to an electron beam and a diffraction pattern is obtained at every orientation (Figure \ref{fig:crystaltilt} (a)). 3D-ED has emerged as a compelling quantitative technique for structure determination in molecular and materials science \citep{palatinus_hydrogen_2017}, particularly for nanocrystalline samples unsuitable for traditional X-ray analysis. The interaction of electrons with matter, which is Coulombic in nature, is orders of magnitude stronger than that of X-rays \citep{henderson1995potential}; this allows high-resolution diffraction patterns to be obtained from $\mu$m or even nm scale crystals \citep{nannenga2014high} and drives pronounced `dynamical' interactions between the incident and diffracted beams, so that the experimentally recorded intensities become exquisitely sensitive to subtle variations in the underlying structure factors \citep{Cowley1957, humphreys_scattering_1979}. Practically, this extreme sensitivity enables the precise location of weakly scattering hydrogen atoms \citep{palatinus_hydrogen_2017}, robust determination of the absolute structure of chiral systems \citep{brazda_electron_2019}, unambiguous discrimination between atomic species possessing nearly identical scattering factors \citep{gemmi_3d_2019} and even information on charge density deformation due to bonding to be obtained \citep{suresh_ionisation_2024}, provided that dynamical calculations are performed and incorporated into the refinement process in order to accurately interpret the experimental data.

For dynamical refinements of 3D-ED data, the Bloch wave formalism \citep{humphreys_scattering_1979, Bethe1928} is generally employed to simulate integrated diffracted intensities as it can readily accommodate arbitrary crystal orientations \citep{palatinus_structure_2015}. While this formalism provides a physically accurate description of multiple elastic scattering, each evaluation of the scattering matrix scales as $\mathcal{O}(N^3)$ and a full 3D-ED refinement, which typically involves $10^3$--$10^5$ evaluations, can quickly scale to hours or even days. Within this already heavy computational budget, a critical challenge is modeling the variation in apparent thickness as a crystal is tilted relative to the electron beam \citep{suresh_ionisation_2024}. Dynamical diffracted intensities are highly sensitive to thickness \citep{zuo_automated_1991}, making accurate thickness modeling essential for quantitative refinement. Real crystals possess irregular shapes, bending, surface roughness, and defects that induce complex, thickness profiles as a function of crystal orientation that may deviate from simple geometric expectations. Current workflows typically define or hand-tune geometric models \citep{palatinus_structure_2015}, which may lack the flexibility to capture experimentally realistic morphologies. Additionally, as these models are not updated within the refinement loop, substantial computational overhead can be expended in search of a correct model. 

In this work, we develop a hybrid physics-machine learning modeling approach to overcome these limitations. By embedding a lightweight neural network that predicts orientation dependent thickness profile directly within a fully differentiable Bloch wave simulator ~\citep{rumelhart_learning_1986, baydin2018automatic} we enable first-order optimizers to jointly update atomic co-ordinates, thermal displacement parameters and the thickness model in a single pass. Our approach represents a significant departure from the second-order methods commonly used in 3D-ED refinements \citep{palatinus_structure_2015, suresh_ionisation_2024, zuo_automated_1991}, which become prohibitively expensive for large parameter spaces or non-linear modules. First-order methods scale efficiently and naturally accommodate neural architectures while maintaining physical interpretability.

We demonstrate that this hybrid differentiable framework matches or exceeds the performance of traditional refinements on both synthetic and experimental 3D‑ED datasets—yielding improved accuracy in atomic positions and thermal parameters. Moreover, because first‑order optimizers incur only linear storage and compute costs as the size of the embedded network grows, we can co‑refine thousands of thickness‑model parameters alongside structural degrees of freedom in a single pass—something that would be infeasible with Hessian‑based, second‑order schemes. While our focus here is on 3D‑ED, the same differentiable‑physics plus learned‑submodel strategy can be applied to other electron microscopy modalities (e.g., 4D‑STEM, convergent‑beam diffraction, scanning‑precession ED) where complex, hard‑to‑model experimental effects likewise limit quantitative analysis.

\section{Results}\label{sec:results}

\subsection{Refinement workflow and thickness optimization}
\label{sec:results-workflow}

\begin{figure}
    \centering
    \includegraphics[width=1.0\linewidth]{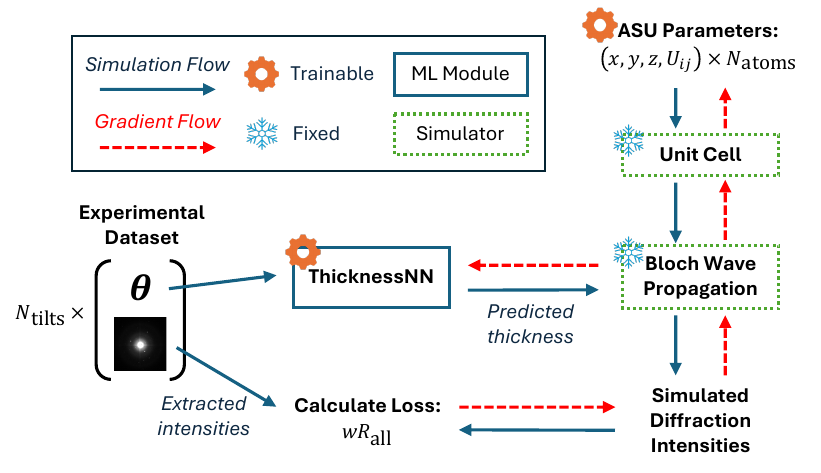}
    \caption{Schematic diagram for the proposed hybrid physics-machine learning model for dynamical electron diffraction. Spatial ($x,y,z$) and thermal displacement ($U_{ij}$) parameters of the atoms in the asymmetric unit, along with neural network parameters of the thickness neural network (ThicknessNN) are jointly optimized using backpropagation using the diffraction loss $wR_{\text{all}}$.}
    \label{fig:workflow}
\end{figure}

Figure \ref{fig:workflow} is a graphical overview of our differentiable refinement workflow. Beginning with the asymmetric unit (ASU), space group symmetry operations are applied to expand the ASU into the unit cell. The unit cell is then rotated to an orientation that matches the experimentally observed diffraction pattern against which we want to compare a simulation. Following this, a Bloch wave calculation is performed to obtain the integrated simulated intensities at this crystal orientation and a loss (in this case a crystallographic R-factor) is calculated between the integrated observed and simulated intensities. Gradients with respect to the R-factor are computed via backpropagation and the ASU parameters, which include atomic positions and isotropic or anisotropic displacement parameters, are updated in order to minimize the loss via gradient descent. The refinement proceeds against integrated intensities and the data is pre-processed according to previously reported virtual frame methods developed by \citet{palatinus_specifics_2019, palatinus_structure_2013, palatinus_structure_2015} and \citet{klar_accurate_2023}. In dynamical electron diffraction refinements, modeling the shape of the crystal and how the apparent thickness along the electron beam varies with the rotation angle is essential for accurately reproducing experimental intensities \citep{palatinus_structure_2015, suresh_ionisation_2024, zuo_automated_1991}. Previous approaches have addressed this challenge by explicitly assuming simple geometric models for the crystal shape—such as a wedge, cylinder, or lens—and analytically deriving the corresponding thickness probability distributions \citep{palatinus_structure_2015}. These models are then adjusted by applying empirical corrections that interpolate between two limiting cases of tilt dependence: an infinite flat plane, where, for tilt angle \(\theta\), the apparent thickness varies as \( t(\theta) = t_0 / \cos\theta \), and a cylindrical or spherical object, where the thickness remains constant, \( t(\theta) = t_0 \). A manually tuned tilt correction parameter is then used to interpolate between these extremes \citep{suresh_ionisation_2024}.

While these fixed models have proved successful, they are limited by a reliance on simplified geometric assumptions and lack flexibility. We instead propose a data-driven approach that leverages the fact that thickness variations and sample geometry are already implicitly encoded in dynamical intensities. By training a neural network to predict thickness as a function of $\theta$ while simultaneously refining the atomic parameters, we should be able to learn the thickness distribution to a good approximation directly from experimental data. This approach enables greater flexibility as it avoids explicit assumptions about crystal geometry while accommodating arbitrary thickness variations. Additionally, due to the differentiable nature of our codebase, the neural network can be easily integrated into our refinement work flow and is compatible with joint optimization of the model alongside the structural parameters of the ASU (Figure \ref{fig:workflow}), resulting in little to no additional computational overhead compared to a refinement of the ASU parameters alone.

\subsection{Refinement on Synthetic Data}
\label{sec:results-synth}

\begin{figure}[ht]
    \centering
    \includegraphics[width=1\linewidth]{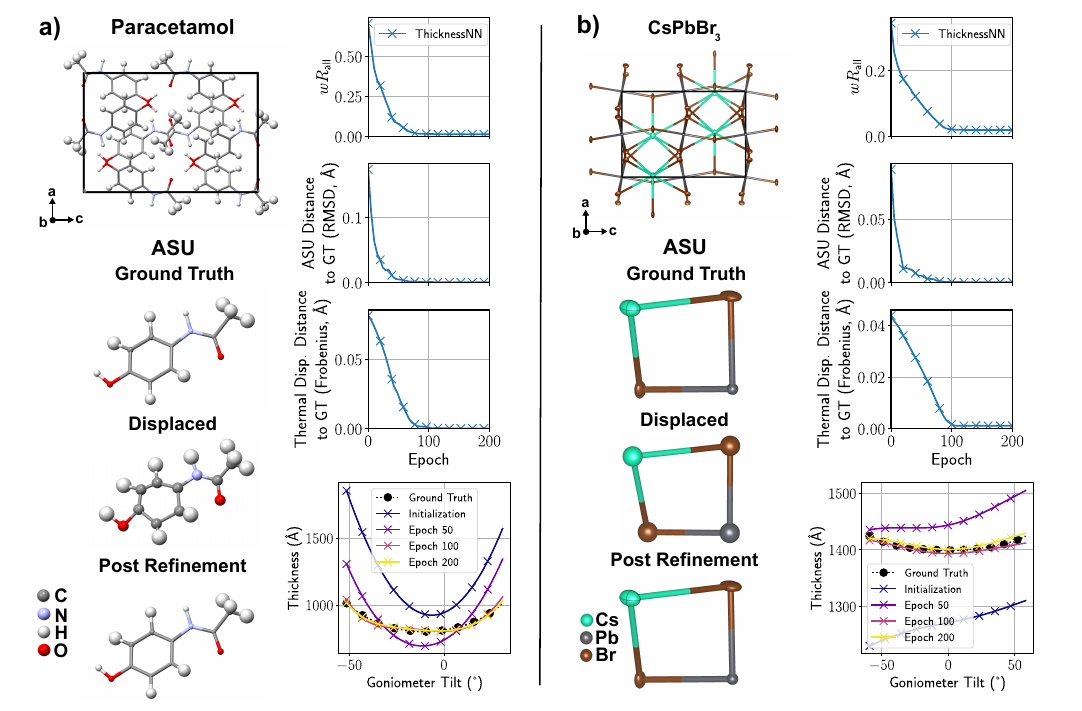}
    \caption{Hybrid Physics-ML model (ThicknessNN) refinements on synthetic Paracetamol and CsPbBr$_3$ 3D-ED datasets. The weighted R-factor (\(wR_{\text{all}}\)), the root mean squared distance (RMSD) to ground truth ASU, the Frobenius norm of the difference in thermal displacements, and the thickness distributions are shown as the refinement progresses from initial 0.2 \AA~randomly displaced starting models. The unit cells, ground truth, starting, and refined models of the ASU are also shown.}
    \label{fig:synth-refine}
\end{figure}

\begin{figure}[ht]
    \centering
    \includegraphics[width=1\linewidth]{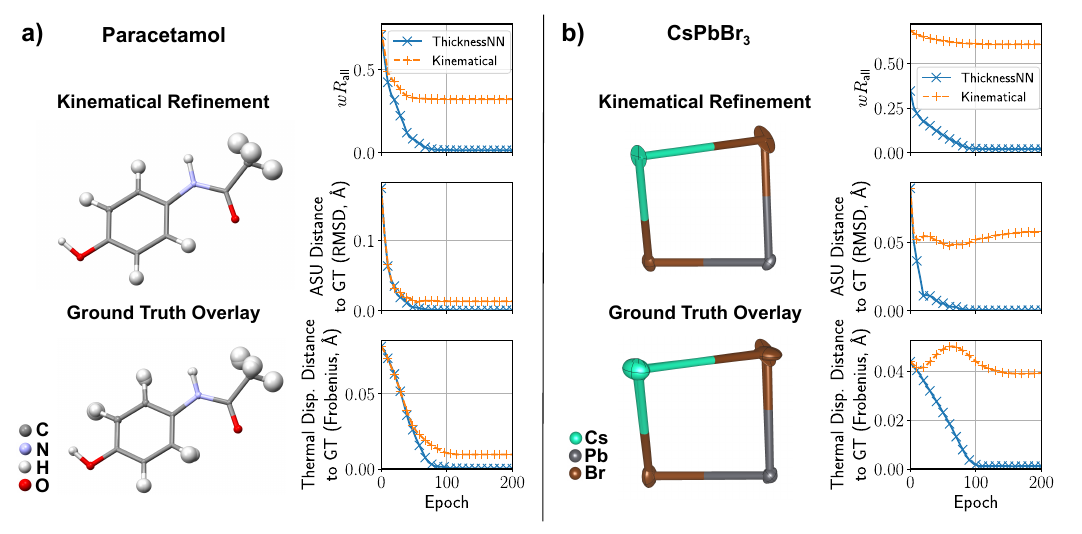}
    \caption{Kinematical refinements on synthetic Paracetamol and CsPbBr$_3$ 3D-ED datasets. The weighted R-factor (\(wR_{\text{all}}\)), the root mean squared distance (RMSD) to ground truth ASU, and the Frobenius norm of the difference in thermal displacements are shown alongside the refined ASU and ground truth overlap from initial 0.2 \AA~randomly displaced starting models. Refinement curves showing the refinement using the proposed hybrid model (ThicknessNN) are also shown for comparison.
    }
    \label{fig:kin-compare}
\end{figure}

To evaluate the ability of our refinement workflow to jointly recover both atomic parameters and sample thickness profiles, we generated synthetic 3D-ED datasets (see Methods) for three representative structures: paracetamol, CsPbBr$_3$, and quartz. Each 3D-ED dataset was corrupted with Poisson noise and was assigned a randomly generated thickness profile that varied smoothly as a function of rotation angle $\theta$, mimicking realistic variations in apparent crystal thickness during data acquisition. 

Initial models for the refinement were created by adding random atomic displacements drawn from a uniform distribution over the range $[-\Delta, \Delta]$~\AA, with maximum displacement magnitude of $\Delta = 0.2$~\AA. Displacements were generated in a symmetry-aware fashion in that atomic positions were perturbed randomly while preserving crystallographic symmetry operations and special Wyckoff positions. Thermal displacement parameters for the starting models were created by scaling the ground truth thermal displacement parameters in proportion to the magnitude of the atomic displacements. Additionally, for CsPbBr$_3$, which has ansiotropic thermal displacement parameters, the starting models were initialized with simplified isotropic values to increase the difficulty of the refinement task.

Figure \ref{fig:synth-refine} shows representative refinements for CsPbBr$_3$ and paracetamol.

We observe that the weighted R-factor (\(wR_{\text{all}}\)), the root mean squared distance (RMSD) to ground truth atomic positions, and the Frobenius norm of the difference in thermal displacement parameters all converge effectively to zero within 200 epochs. These results are promising for several reasons. Firstly, a 0.2~\AA~displacement represents a relatively poor initial model for a crystallographic refinement and we do not employ any stereochemical restraints, yet convergence is still achieved. Secondly, in the cases of CsPbBr$_3$, the refinement successfully recovers anisotropic thermal displacements even when initialized with isotropic ones with a large displacement. Thirdly, the ground truth thickness distribution is successfully recovered alongside the atomic parameters and this remarkable performance continues even when the thickness profile has a complex dependence on $\theta$ (Supplementary Note 1). All of this is achieved using the Adam optimizer \citep{KingmaB14} with default settings and without any hyperparameter tuning. Similar results are observed for the quartz datasets (Table \ref{tab:synth-results} and Supplementary Figure~\ref{supp-fig:synth-refine}), and for models initialised with displacements of 0.1~\AA~(Supplementary Figure~\ref{supp-fig:synth-refine-01}) and 0.3~\AA~(Supplementary Figure~\ref{supp-fig:synth-refine-03}).

To understand the importance of utilising a fully dynamical forward model in the refinement of MicroED data, we compare our results to those obtained with a kinematical model which we approximate by fixing the sample thickness to 10~\AA~such that multiple scattering is negligible in Figure~\ref{fig:kin-compare}. As expected, a degradation in refinement quality is observed when using a kinematical model and it is most pronounced when strong dynamical scattering is present in the dataset. For instance, in CsPbBr$_3$, where substantial dynamical effects are expected due to the combination of the presence of heavy elements (Cs and Pb) and a large simulated crystal thickness (Figure \ref{fig:synth-refine}), the final RMSD in atomic positions and the Frobenius norm of thermal displacement deviations remain substantially large compared to the dynamical models. 
In contrast, the performance gap is smaller for paracetamol, which consists of lighter atoms and was simulated with a thinner crystal.

Summary metrics for the converged refinements across all three structures are shown in Table \ref{tab:synth-results}. Here we also ablate performance against a dynamical refinement with an imperfect thickness model, where we assume a constant effective thickness for the entire tilt series rather than a detailed thickness profile. The recovered single thickness values along with the ground truth thickness distributions are shown in Supplementary Figure \ref{supp-fig:thicknesses}. We find that the single-thickness model also consistently under-performs the hybrid approach. This is especially pronounced in quartz where strong variation with $\theta$ is present. We observe similar trends for various levels of structural perturbation. Full refinement curves and results across datasets are provided in Supplementary Figures \ref{supp-fig:synth-refine-01}, \ref{supp-fig:synth-refine} and \ref{supp-fig:synth-refine-03}. We further note runtime comparisons to second order methods in Supplementary Note 2. 

\begin{table}[h]
\caption{Results comparing refinements with different modeling assumptions (kinematical, constant thickness dynamical, and the ThicknessNN hybrid model) on synthetic 3D-ED datasets. We report the weighted R-factor (\(wR_{\text{all}}\)), the root mean squared distance (RMSD) to ground truth ASU, and the Frobenius norm of the difference in thermal displacements for converged models after refining for 200 epochs from an initial 0.2 \AA~randomly displaced starting model.}
\label{tab:synth-results}
\begin{tabular}{@{}p{1.8cm}p{2.4cm}p{1.8cm}p{1.8cm}p{2.3cm}@{}}
\toprule
\textbf{Sample}      & \textbf{Metric}                    & \textbf{Kinematical} & \textbf{Dynamical (Single Thickness)} & \textbf{Hybrid Physics-ML (ThicknessNN)} \\ \hline
\textbf{Quartz}      & $wR_{\text{all}}$                  & 0.452                & 0.207                               & \textbf{0.012   }                                 \\
                     & ASU RMSD (Å)                       & 0.0347               & 0.0026                              & \textbf{0.0003 }                                  \\
                     & Thermal Disp. (Å) & 0.0154               & 0.0023                              & \textbf{0.0001 }                                  \\ \hline
\textbf{CsPbBr$_3$}  & $wR_{\text{all}}$                  & 0.607                & 0.121                               & \textbf{0.020 }                                   \\
                     & ASU RMSD (Å)                       & 0.0575               & 0.0080                              &\textbf{0.0004}                                   \\
                     & Thermal Disp. (Å) & 0.0392               & 0.0160                              & \textbf{0.0011 }                                  \\ \hline
\textbf{Paracetamol} & $wR_{\text{all}}$                  & 0.323                & 0.0279                              & \textbf{0.014}                                    \\
                     & ASU RMSD (Å)                       & 0.0136               & 0.0007                              &\textbf{0.0006  }                                 \\
                     & Thermal Disp. (Å) & 0.0096               & 0.0009                              & \textbf{0.0002 }  \\ 
\bottomrule
\end{tabular}
\end{table}

\subsection{Refinement on Experimental Data}
\label{sec:results-exp}

\begin{table}[h]
\caption{Results of dynamical refinements performed on experimental data. *$N_{\text{obs}}, N_{\text{all}}$ for the paracetamol dataset were not explicitly reported at 1.2 $(\text{Å}^{-1})$ resolution. These values were estimated by examining the number of reflections that passed $R_{\text{obs}}$ and $wR_{\text{all}}$ filters in the experimental dataset provided in Ref. \citep{noauthor_pets_nodate}}
\label{tab:experimental_refinement_results}
\begin{tabular}{@{}p{1.8cm}p{2cm}p{1.8cm}p{2cm}@{}}
\toprule
\textbf{Sample}  & \textbf{Metric} & \textbf{DYNGO-JANA} & \textbf{Hybrid Physics-ML} \\ \midrule
\textbf{Quartz} & $R_{\text{obs}}$ &5.7 \% \citep{klar_accurate_2023}    &   4.2 \%  \\
(\textit{Continuous-} & $wR_{\text{all}}$    &  6.6 \% \citep{klar_accurate_2023}   &  4.3 \%   \\
\textit{Rotation}) & $N_{\text{obs}}, N_{\text{all}}$  &994, 1710  & 957, 1734  \\ 
& $g_{\text{max}} (\text{Å}^{-1})$  & 1.6   &   1.6 \\ \midrule
\textbf{CsPbBr$_3$} & $R_{\text{obs}}$ & 5.5 \% \citep{suresh_ionisation_2024}  & 6.4   \%     \\
(\textit{Continuous-}& $wR_{\text{all}}$    & 6.8 \% \citep{suresh_ionisation_2024}   & 6.7 \%   \\ 
\textit{Rotation}) & $N_{\text{obs}}, N_{\text{all}}$  & 2709, 16345  &  2714, 17095 \\ 
& $g_{\text{max}} (\text{Å}^{-1})$   & 2.0   &   2.0 \\ \midrule
\textbf{Paracetamol} & $R_{\text{obs}}$ &  9.2 \% \citep{noauthor_pets_nodate}  & 10.15 \%       \\
(\textit{Precession}) & $wR_{\text{all}}$    & 10.34 \% \citep{noauthor_pets_nodate}  &   8.4 \% \\
& $N_{\text{obs}}, N_{\text{all}}$  & 2738*, 6669* & 2697, 6621  \\ 
& $g_{\text{max}} (\text{Å}^{-1})$   & 1.2   &  1.2  \\

\bottomrule
\end{tabular}
\end{table}

In the refinement of synthetic datasets, we assume purely elastic scattering with added Poisson noise, and the experimental orientation is considered to be perfectly known. In contrast, real experimental data are affected by additional complexities, including inelastic scattering processes—such as plasmon losses and thermal diffuse scattering \citep{mendis_modelling_2024, mendis_phase_2023}—as well as detector noise \citep{palatinus_specifics_2019}. Accurate frame-by-frame orientation determination is also essential to match simulations to measured intensities \citep{palatinus_specifics_2019}. To evaluate the robustness of our proposed hybrid physics–ML refinement approach on real data, we reproduced dynamical refinements for three previously published 3D electron diffraction (3D-ED) datasets: quartz \citep{klar_accurate_2023}, CsPbBr$_3$ \citep{suresh_ionisation_2024}, and paracetamol \citep{noauthor_pets_nodate}. In each case, the original analyses involved data processing using the PETS2 software package \citep{palatinus_specifics_2019}, followed by dynamical refinements with DYNGO and JANA.

For our refinements, we used the same PETS2 output files reported in the literature but replaced the DYNGO-JANA pipeline with our own hybrid physics-ML refinement code. To our knowledge, this is the first demonstration and benchmarking of an alternative codebase to JANA and DYNGO for dynamical refinement of 3D-ED datasets. Table~\ref{tab:experimental_refinement_results} summarizes our results. Across all datasets, the number of $N_{\text{obs}}$ (I $>$ 3$\sigma$) and $N_{\text{all}}$ intensities compares very closely between our refinements and the original studies. The small differences in $N_{\text{obs}}$ and $N_{\text{all}}$ between our results and those in the literature likely arise from variations in orientation search protocols and parameterizations. For continuous rotation datasets (quartz and CsPbBr$_3$), we perform a three-parameter simplex search involving rotations about the crystal’s  $x$, $y$, and $z$ axes (see Methods). In contrast, published studies appear to predominantly use a two-parameter Euler angle approach — $\phi$  (rotation about the laboratory z-axis) and $\theta$ (tilt about the new x-axis) — for both continuous rotation and precession datasets \citep{klar_accurate_2023, palatinus_structure_2013}. For precession electron diffraction data (Paracetamol), we follow this convention, where $\phi$ defines the tilt direction and $\theta$ its amplitude.

\begin{figure}
    \centering
    \includegraphics[width=0.5\linewidth]{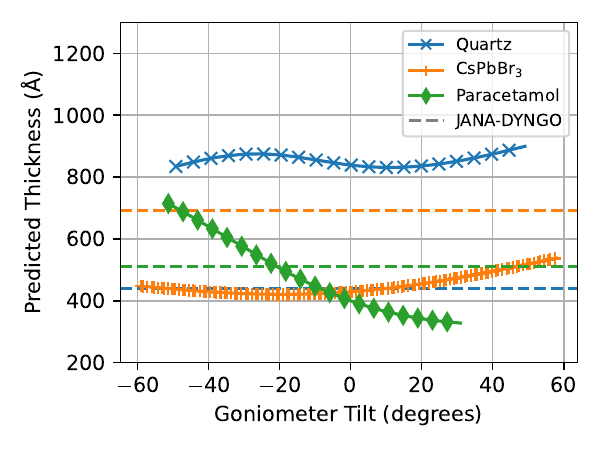}
    \caption{Experimental thickness curves recovered from the hybrid physics-ML model and DYNGO-JANA (dashed lines) for Quartz, CsPbBr$_3$, and Paracetamol datasets}
    \label{fig:experimental_thicknesses}
\end{figure}

For the quartz dataset, the hybrid physics-ML method achieves a lower $R_{\text{obs}}$ value (4.2 \%) and $wR_{\text{all}}$ value (4.3 \%) compared to the DYNGO-JANA values of (5.7 \%) and  (6.6 \%), indicating improved fit to the observed intensities. This is likely due to the thickness modeling employed in the hybrid physics-ML refinement; in Ref. \citep{klar_accurate_2023}, from which the quartz dataset is obtained, there is no explicit mention of modeling the thickness of the crystal as a function of $\theta$ beyond a single refined thickness value of $\sim$ 44 nm. By comparison, the ThicknessNN utilised in the hybrid physics-ML method recovers a more complex relationship of apparent thickness as a function of $\theta$ in addition to an overall larger crystal thickness at each orientation ($\sim$ 85 nm average). The recovered thickness distribution (Figure ~\ref{fig:experimental_thicknesses}) is plausible given the complex wedge like geometry of the quartz crystal and the substantial specimen movement during data acquisition (See Supplementary Video 1), though we note that it is difficult to derive precise specimen geometries from experimental electron microscopy images.

For CsPbBr$_3$, the recovered $R_{\text{obs}}$ and $wR_{\text{all}}$ values are comparable between our hybrid Physics-ML model (6.4\% $R_{\text{obs}}$, 6.7\% $wR_{\text{all}}$) and those reported in the original study (5.5\% $R_{\text{obs}}$, 6.8\% $wR_{\text{all}}$) \citep{suresh_ionisation_2024}. These similar residuals are expected given that both approaches incorporate an explicit thickness model. However, the underlying recovered thickness distributions differ. The original study reports a constant thickness of 69.1 nm across the dataset and noted that introducing a tilt correction parameter—intended to vary thickness as a function of $\theta$ — did not improve the residuals. This suggests a sample geometry approximating a uniform cylindrical rod, such that $t(\theta) = t_0$. In contrast, our model recovers a thickness distribution with a mean of 44.0 nm and a mild dependence on $\theta$ (see Figure~\ref{fig:experimental_thicknesses}). An advantage of our approach is that we do not require an explicit tilt correction parameter, which in the original study may have necessitated multiple refinement cycles to identify an optimal form. In this case, both models yield similar residuals because the thickness distribution is relatively simple. However, in more complex scenarios—such as the quartz dataset discussed earlier—a manually defined tilt correction function would likely fail to capture the true variation, highlighting the flexibility and generality of our approach.

For the paracetamol dataset, our $R_{\text{obs}}$ is higher than reported (10.15\% vs 9.2\%), but our $wR_{\text{all}}$ is lower (8.4\% vs 10.34\%), which is atypical, but has been observed in other dynamical refinements that utilise the same weighting in the calculation of $wR_{\text{all}}$ (see Methods) \citep{palatinus_structure_2013}. The recovered thickness distribution with the hybrid physics-ML model is asymmetric, which is roughly consistent with the rotation range covered by the goniometer during the experiments (-52$\degree$ to +30$\degree$). Interestingly, the mean value of thickness distribution recovered from the hybrid physics-ML model (47.5 nm) compares closely to the single thickness value refined with the DYNGO-JANA model (51.1 nm~\citep{noauthor_pets_nodate}, Figure~\ref{fig:experimental_thicknesses}).

One of the key advantages of electron diffraction is its high sensitivity to the chirality (handedness) of non-centrosymmetric crystal structures. In such systems, chirality is encoded in the violation of Friedel’s Law, which results in intensity differences between symmetry-related reflections \( \mathbf{g} \) and \( -\mathbf{g} \). This asymmetry arises from multiple scattering effects and cannot be interpreted using the kinematical approximation alone. Instead, full dynamical simulations are required, involving refinements of both enantiomorphic models to determine the correct absolute structure \citep{brazda_electron_2019, klar_accurate_2023}. To verify that our hybrid physics-ML model can recover these differences we perform independent refinements of both enantiomers (\textit{P}3$_2$21   and  \textit{P}3$_1$21) of quartz to replicate the results presented in \citep{klar_accurate_2023}. We obtain substantially lower residual factors with \textit{P}3$_2$21 enantiomer (Table \ref{tab:enant}), which is consistent with previous results \citep{klar_accurate_2023} and demonstrates that the hybrid physics-ML model is capable of accurately determining the absolute structure of chiral systems.

\begin{table}[h]
\caption{Absolute structure determination from the 3D-ED quartz dataset.}
\label{tab:enant}
\begin{tabular}{@{}p{2.0cm}p{3.2cm}p{3.2cm}@{}}
\toprule
\textbf{Sample}  & \textbf{Enantiomorph 1} & \textbf{Enantiomorph 2} \\ \midrule
\textbf{Quartz} &  &   \\\midrule
 Spacegroup & \textit{P}3$_2$21    &  \textit{P}3$_1$21  \\\midrule
$g_{\text{max}} (Å^{-1})$ &  1.6 & 1.6   \\ \midrule
 $N_{\text{obs}}, N_{\text{all}}$ & 957, 1734 & 957, 1734  \\ \midrule
$wR_{\text{all}}$ & \textbf{4.3 \%}    & 7.1 \%  \\ \midrule
 $R_{\text{obs}}$ & \textbf{4.2 \%}  & 6.2 \% \\ \midrule

\bottomrule
\end{tabular}
\end{table}

\section{Discussion}\label{sec:disc}

Our results demonstrate that hybrid physics–machine learning models, built on a foundation of differentiable physics, offer a powerful and flexible framework for dynamical refinement of 3D electron diffraction data. By combining a fully differentiable Bloch wave simulator with neural networks, we enable direct learning of complex, non-convex thickness distributions from experimental data—without relying on geometric assumptions or empirical correction schemes that underpin traditional approaches. 

This capacity to extract latent physical variables directly from intensities opens the door to modeling other challenging experimental effects, such as inelastic scattering, which have historically resisted tractable physical modeling. The modularity and scalability of our implementation in PyTorch \citep{PaszkeGMLBCKLGA19} make such extensions natural, and suggest broad applicability beyond 3D-ED to other electron microscopy modalities where hybrid physics–ML models may similarly enhance accuracy, interpretability, and computational efficiency. We anticipate that such models will become increasingly central to quantitative electron microscopy, providing a new foundation for interpreting experimental results that remains faithful to physics while flexibly adapting to experimental complexity.

\section{Methods}
\label{sec:methods}

\subsection{Hybrid ML-Physics Model}
\label{sec:methods-model}

Our forward diffraction model, shown in Figure \ref{fig:workflow}, consists of a hybrid neural-network based thickness model and a differentiable physics simulator, combining the physical accuracy of the Bloch wave formalism and the flexibility of neural network function approximation. 

\subsubsection{Differentiable Bloch Wave Propagation}
\label{sec:diffbloch}

The Bloch wave method, developed by Bethe \cite{Bethe1928} and later expanded by others \cite{humphreys_scattering_1979, hirsch1965electron, Cowley1957}, models elastic electron scattering in perfect crystals across arbitrary orientations and thicknesses, yielding diffracted intensities that can be directly compared to 3D-ED data. Its matrix formulation is naturally differentiable, making it well adapted to gradient-based optimization in dynamical refinement \cite{palatinus_structure_2013}. The Bloch wave formalism expands the wavefunction of the incident electron \( \psi(\mathbf{r}) \) into a finite set of plane-wave states permitted by the periodic crystal potential \( V(\mathbf{r}) \). For a complete theoretical treatment, see Spence and Zuo~\cite{Spence2020}. The periodic crystal potential is expressed as $V(\mathbf{r}) = \sum_{\mathbf{g}} V_{\mathbf{g}} \mathrm{e}^{2\pi i \mathbf{g} \cdot \mathbf{r}}$,
where each reciprocal lattice vector \( \mathbf{g} \) corresponds to a Fourier component \( V_{\mathbf{g}} \), determined by the underlying atomic model. 

To compute \( V_{\mathbf{g}} \), the ASU is expanded to the full unit cell using the crystal's space group operations. Each atom $i$ contributes an electron scattering factor \( f_i^e \), evaluated at scattering vector \( s = |\mathbf{\frac{g}{2}}| \), modulated by the Debye–Waller factor  \( B_i\) that accounts for atomic thermal motion. $V_{\mathbf{g}}$ then takes the form:
\[
V_{\mathbf{g}} = \frac{1}{\Omega} \sum_i f_i^e(\mathbf{s}) \, \exp\left( -B_i s^2 \right) \, \exp(-2\pi i \mathbf{g} \cdot \mathbf{r}_i),
\]
where \( \Omega \) is the unit cell volume and \( \mathbf{r}_i \) is the atomic position. Thermal vibrations reduce the sharpness of the periodic potential, leading to overall damping of diffracted intensities, with stronger attenuation at higher scattering angles. For isotropic displacements, the Debye–Waller factor is given by \( B_i = 8\pi^2 \langle u_i^2 \rangle \), where \( \langle u_i^2 \rangle \) is the Cartesian atom-specific mean-square displacement. In the anisotropic case, thermal motion is described by a symmetric second-rank tensor \( U_{ij} \), defined as the covariance \( U_{ij} = \langle u_i u_j \rangle \) of atomic displacements along Cartesian directions \( i \) and \( j \) \citep{GrosseKunstleve2002}.  To express \( U_{ij} \) in crystallographic fractional coordinates, it is transformed using the orthogonalization matrix \( A \):
\[
U_{ij}^\star = A^{-1} U_{ij}^{\mathrm{cart}} A^{-\mathsf{T}}
\]
This transformation ensures compatibility with space group symmetry operations. To maintain physical validity and enable stable refinement, the tensor is parameterized via its Cholesky decomposition $U = L L^\top$, where \( L \) is a lower triangular matrix. This guarantees positive-definiteness and enables smooth optimization. During asymmetric unit expansion, each \( U_{ij}^\star \) is rotated under the space group operations and then converted back to Cartesian form for use in scattering calculations.

\subsubsection{Apparent Thickness Neural Network}
\label{sec:methods-thickness}

To model orientation-dependent variations in apparent crystal thickness, we implemented a lightweight neural network module (\texttt{ThicknessNN}) in PyTorch \citep{PaszkeGMLBCKLGA19} that maps a rotation angle $\theta$ to a predicted mean thickness $\mu$ and associated standard deviation $\sigma$. The model consists of a feedforward architecture with two hidden layers of 64 units each and \texttt{Tanh} activation functions (to promote smoothness in $\theta$), followed by a two-node output layer that returns $\mu$ and $\log \sigma^2$. The predicted standard deviation is scaled to lie within a user-defined range to ensure numerical stability. Often the user will have intuitive knowledge of both the thickness range (e.g. 10 - 200 nm) and the shape of its distribution. These priors can be flexibly incorporated into the ThicknessNN to warm-start the optimization. For example, we incorporate an optional quadratic prior by expressing $\mu(\theta)$ as a weighted combination of a learnable quadratic function $a\theta^2 + b\theta + c$ (with $a > 0$ enforced via exponential parameterization) and a residual term learned from the network. This hybrid formulation allows the model to learn deviations from simple geometric thickness profiles while retaining regularizing inductive structure. In synthetic data experiments, we utilized this prior for quartz and paracetamol, however we soon realised that it was not required for reliable thickness distribution recovery. For subsequent experiments including synthetic CsPbBr$_3$ and experimental data, we did not add this prior, though we note in particular high-noise or data scarce environments (such as refining on partial datasets), these priors may prove useful.

Both the predicted mean and standard deviation can be used to generate thickness samples via the reparameterization trick, analogous to variational autoencoders (VAEs) \citep{KingmaW13}. These sampled thicknesses can be used to perform multiple Bloch wave simulations per orientation, with the resulting intensities averaged to marginalize over thickness uncertainty. However, we found that this approach had minimal effect on the recovered residual R-factors. Consequently, for all refinements presented in this work, we used the predicted mean thickness alone to perform a single Bloch wave simulation per orientation.

\subsection{Refinements, loss functions and reflection filtering}
\label{sec:methods-loss-funcs}
 For calculation of residuals $wR_{\text{all}}$ and $R_{\text{obs}}$ we followed the conventions and definitions outlined in \citet{klar_accurate_2023}.

\begin{align}
R &= \frac{\sum \left| \sqrt{I_{\text{obs}}} - \sqrt{I_{\text{calc}}} \right|}{\sum \sqrt{I_{\text{obs}}}} \\
wR &= \sqrt{ \frac{\sum \left( w \left| I_{\text{obs}} - I_{\text{calc}} \right| \right)^2 }{ \sum \left( w I_{\text{obs}} \right)^2 } } \\
w &= \left( \sigma\left( \sqrt{I_{\text{obs}}} \right)^2 + \left( u \sqrt{I_{\text{obs}}} \right)^2 \right)^{-1/2}
\end{align}
Where the sums run over all reflections for the calculation of $wR_{\text{all}}$, and only over observed reflections with $I_{\text{obs}} > 3\sigma(I_{\text{obs}})$ for $R_{\text{obs}}$. The instability factor $u$ was set to 0.01. Reflections with either weak or negative intensities, with $I_{\text{obs}} < 0.01\,\sigma(I_{\text{obs}})$ were set to 0, and their uncertainties were adjusted such that
\[
\sigma\left( \sqrt{I_{\text{obs}}} \right) = 5 \sqrt{ \sigma(I_{\text{obs}}) }.
\]

All refinements of atomic positions, anisotropic thermal displacement, ThicknessNN weights and biases and orientations proceeded against $wR_{\text{all}}$ and did not utilise any restraints or constraints beyond those imposed by crystal symmetry for a particular spacegroup. For all experiments, we refined parameters using the Adam optimizer~\citep{KingmaB14} over 200 epochs (passes over the full dataset of rotations) using a full batch gradient descent (we average the loss over all rotations for each gradient step), and a learning rate of 0.001 for all parameters.

For selection of integrated intensities, the methodologies to calculate $S_g$, $S_{g}^{max}$, $D_{sg}$ and $R_{sg}$ described in Ref. \citep{klar_accurate_2023} were again followed. Values for $S_g$, $S_{g}^{max}$, $D_{sg}$ and $R_{sg}$ utilized in the refinements of experimental data were as previously reported \citep{klar_accurate_2023, suresh_ionisation_2024, noauthor_pets_nodate}.

\subsection{Orientation refinement}
Orientation refinement for each rotation was performed using either a three-parameter Nelder–Mead simplex search or a modified simplex approach based on the methodology described by \citet{palatinus_structure_2013}. In the Nelder–Mead approach, the crystal orientation was parameterized as a product of three small-angle rotations about the crystal’s \(x\), \(y\), and \(z\) axes, with the optimization objective being the minimization of the residual \(wR_{\text{all}}\). The three rotation angles \(\alpha\), \(\beta\), and \(\omega\) were optimized using the \texttt{scipy.optimize.minimize} function with \texttt{method='Nelder-Mead'}, initialized from a regular simplex \citep{2020SciPy-NMeth}. At each optimization step, an updated orientation matrix was constructed and used in the forward simulation of dynamical diffraction intensities. The updated orientation matrix had the form
\[
R = R_z(\omega) R_y(\beta) R_x(\alpha)
\]

The alternative two parameter orientation search followed a method whereby a discrete hexagonal scan over orientations defined by two angular parameters: $(\phi)$, a rotation about the laboratory $z$-axis, and $\theta$, a tilt about the rotated \(x\)-axis was performed. Trial orientations were generated by rotating the current orientation matrix as
\[
R_{\text{new}} = R_z(\phi) R_x(\theta) R_z(-\phi) R_{\text{current}},
\]
with $(\phi)$ varied in 60$\degree$ increments and $\theta$ halved iteratively if no improvement in \(wR_{\text{all}}\) was found. This approach effectively performs a simplex-like local search over a cone of directions around the current orientation.

Both search protocols used the full experimental–simulated diffraction matching pipeline described above, including filtering of reflections and scaling factor optimization, to evaluate the objective function at each orientation step. 

\subsection{Synthetic Data Generation and Noise Addition}

The synthetic data utilised in this study was generated from atomic co-ordinates obtained from successful experimental dynamical refinements of quartz \citep{palatinus_hydrogen_2017, suresh_ionisation_2024}.

These three structures were chosen as they span many of the current application areas of dynamical refinements of 3D-ED data (inorganic materials science and small molecule crystallography) and will exhibit varying degrees of dynamical effects (we expect CsPbBr$_3$ to have more dynamical scattering than paracetamol for instance).

To model experimental noise in simulated diffraction data, we applied Poisson-distributed noise across the full dataset. All simulated intensities were first linearly scaled such that the brightest reflection across the dataset matched a predefined dynamic range maximum (e.g. $I_{\max}$ = $10^6$ or $I_{\max}$ = $10^2$ etc). This ensures control over the signal-to-noise ratio (SNR), with lower dynamic range values producing proportionally higher noise relative to signal, thereby emulating low-dose data acquisition conditions. After scaling, reflections were removed from each individual pattern if their intensities were more than a factor of the dynamic range below the global maximum, i.e., if their scaled intensity fell below one count. This step mimics a practical detector noise floor and enforces the selected dynamic range as a hard cutoff on observable intensities. Poisson noise was then applied to the remaining scaled intensities using the \texttt{torch.poisson} function from the PyTorch library, which samples from a Poisson distribution with mean equal to the scaled intensity. This implementation captures counting statistics typical of electron diffraction experiments while preserving the statistical structure of the dataset. The synthetic data presented in the main text was generated using a dynamic range of $10^6$, consistent with the capabilities of state-of-the-art photon-counting detectors. The methodology was also evaluated under high-noise conditions using a reduced dynamic range of $10^2$ to assess robustness in low-SNR regimes (Supplementary Figure~\ref{supp-fig:high_noise_thickness}). Uncertainties for each reflection were estimated as the square root of the noisy intensities, i.e., $\sigma_i = \sqrt{I_i}$, consistent with Poisson counting statistics.

\backmatter

\bmhead{Acknowledgements}

TASD acknowledges the support of a Schmidt Science Fellowship and an Oppenheimer Research Fellowship. SM acknowledges funding from EPSRC Centre for Doctoral Training in Autonomous Intelligent Machines and Systems (Grant No: EP/S024050/1). BC acknowledges funding from Queens' College Cambridge and the Stamps Scholars Program. YG is supported by a Turing AI Fellowship financed by the UK government’s Office for Artificial Intelligence, through UK Research and Innovation (grant reference EP/V030302/1) and delivered by the Alan Turing Institute.

\noindent\textbf{Competing interests} \\
The authors declare no competing interests.

\vspace{6pt}

\noindent\textbf{Data availability} \\
All relevant data supporting the findings of this study are available from the corresponding author upon request. Data will be made publicly available in a suitable repository upon publication.

\vspace{6pt}

\noindent\textbf{Code availability} \\
The custom code used in this study is not currently publicly available as it is undergoing preparation for stable open-source release in collaboration with software developers. The code will be made openly accessible upon publication, and a link to the repository will be provided at that time. In the interim, the code can be shared with reviewers or editors upon request for the purpose of peer review.

\vspace{6pt}

\noindent\textbf{Author contributions} \\
\textbf{Conceptualisation:} TASD, PAM, YG. \textbf{Methodology:} TASD, SM, PAM, YG \textbf{Investigation:} TASD, SM, BC. \textbf{Data Curation:} SM, TASD. \textbf{Formal Analysis:} SM, TASD \textbf{Software:} SM, TASD, BC. \textbf{Writing - original draft:} SM, TASD, BC. \textbf{Writing - review and editing:} All authors. \textbf{Supervision:} TASD, PAM, YG, SR. \textbf{Funding Acquisition:} TASD, PAM, YG.


\bibliography{sn-bibliography}

\end{document}


\title[Supplementary Information]{Supplementary Information for: Hybrid Physics-Machine Learning Models for Quantitative Electron Diffraction Refinements}

\author[1]{\fnm{Shreshth A.} \sur{Malik}}
\equalcont{These authors contributed equally to this work.}

\author*[1,2]{\fnm{Tiarnan A.S.} \sur{Doherty}}\email{td404@cam.ac.uk}
\equalcont{These authors contributed equally to this work.}

\author[2]{\fnm{Benjamin} \sur{Colmey}}

\author[3]{\fnm{Stephen J.} \sur{Roberts}}

\author*[1]{\fnm{Yarin} \sur{Gal}}\email{yarin.gal@cs.ox.ac.uk}

\author*[2]{\fnm{Paul A.} \sur{Midgley}}\email{pam33@cam.ac.uk}

\affil[1]{\orgdiv{OATML, Department of Computer Science}, \orgname{University of Oxford}, \orgaddress{\street{Wolfson Building, Parks Rd}, \city{Oxford}, \postcode{OX1 3QG},  \country{United Kingdom}}}

\affil[2]{\orgdiv{Department of Materials Science and Metallurgy}, \orgname{University of Cambridge}, \orgaddress{\street{27 Charles Babbage Rd}, \city{Cambridge}, \postcode{CB3 0FS}, \country{United Kingdom}}}

\affil[3]{\orgdiv{MLRG, Department of Engineering Science}, \orgname{University of Oxford}, \orgaddress{\street{Eagle House, Walton Well Road}, \city{Oxford}, \postcode{OX2 6ED},  \country{United Kingdom}}}

\maketitle

\section*{Supplementary Note 1}
\label{sec:supp-note-1}
\begin{figure}[H]
    \centering
    \includegraphics[width=1\linewidth]{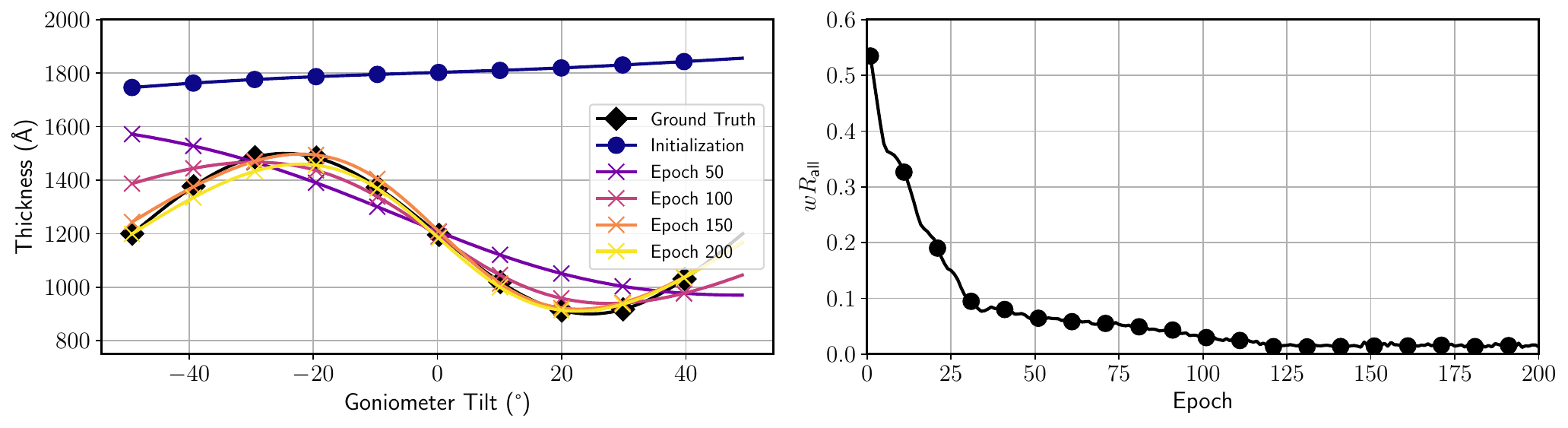}
    \caption{Refinement of a sinusoidal synthetic thickness distribution using the thickness neural network. The network recovers the true underlying thickness distribution (left) via minimizing the diffraction loss (right).}
    \label{fig:thicknessnn}
\end{figure}

To demonstrate the recovery of arbitrary thickness distributions, we test our method on a synthetic 3D-ED dataset of a quartz crystal corrupted with Poisson noise (see Methods). The ground truth thickness distribution varies non-linearly with $\theta$ in a sine-like fashion. Remarkably, over 200 epochs of training, the ThicknessNN recovers the full underlying distribution directly from the data, without relying on any prior assumptions about the sample geometry (Figure \ref{fig:thicknessnn}). The only inductive bias imposed, in this case, is a smoothness constraint through the use of Tanh activations, allowing the model to flexibly learn arbitrary variations as a function of $\theta$ (see Methods). This highlights the model’s capacity to fit complex, non-convex thickness profiles purely through gradient-based refinement. By contrast, existing parametric models, which seem restricted to recover convex functions (e.g., cosine-like) that are symmetric about $\theta = 0$ or assume constant thickness \citep{suresh_ionisation_2024}, would not be able to recover this thickness distribution. 

\begin{figure}[H]
    \centering
    \includegraphics[width=1\linewidth]{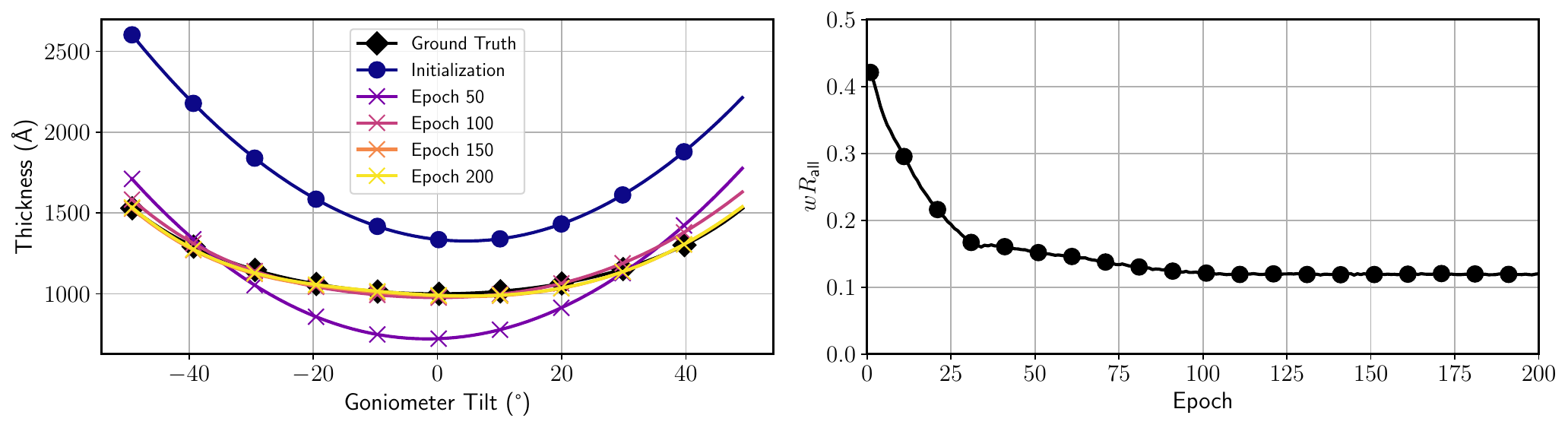}
    \caption{Successful thickness recovery in a quartz dataset simulated with high noise. The total dynamic range of the diffraction patterns was set to $10^2$.}
    \label{fig:high_noise_thickness}
\end{figure}

We find that our method also accurately recovers thickness distributions in high noise limits (Figure \ref{fig:high_noise_thickness}). 

\section*{Supplementary Note 2}
\label{sec:supp-note-2}

\begin{figure}[H]
    \centering
    \includegraphics[width=1\linewidth]{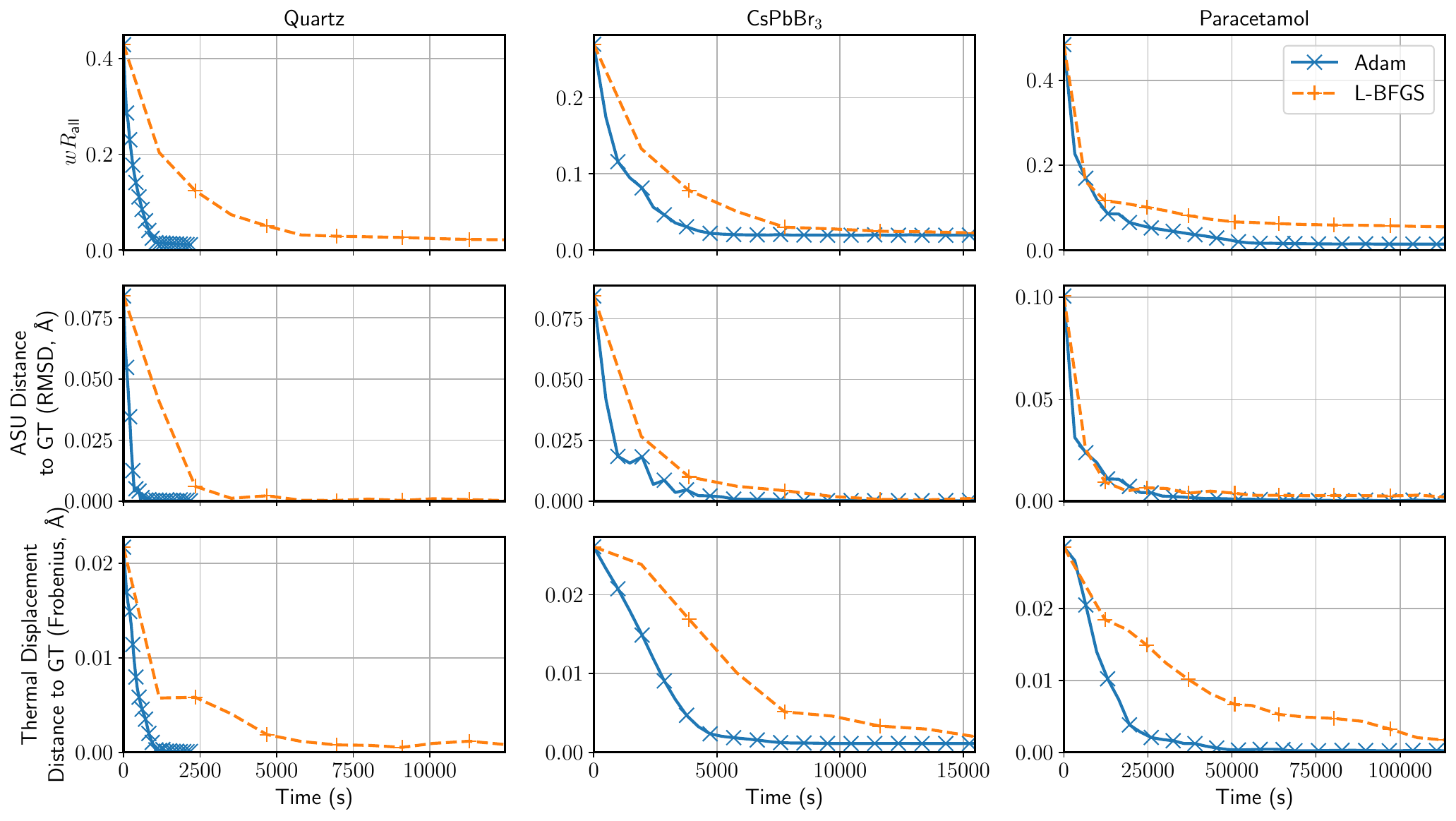}
    \caption{Refinement of synthetic crystal structures starting from 0.1 Å random maximum displacement from the ground truth, comparing Adam and L-BFGS optimization. The mean diffraction intensity loss across all rotations in the dataset $wR_{\text{all}}$, the root mean squared displacement (RMSD) to the unperturbed asymmetric unit, and the Frobenium norm of difference in thermal displacement parameters are shown for each material as a function of the refinement runtime.}
    \label{fig:synth-refine-adam-lbfgs}
\end{figure}

To assess the efficiency of the first-order optimization approach utilizing the Adam optimizer \citep{KingmaB14}, which is presented in the main text, we compared runtimes against second-order optimizers such as L-BFGS~\citep{LiuN89}. The results are summarized in Figure \ref{fig:synth-refine-adam-lbfgs}. For example, on the quartz structure with a 0.1 Å displacement, the refinement using Adam achieves comparable or better accuracy with $\sim50\times$ speed-up to reach a $wR_{\text{all}}$ of 0.02. While the speed improvements on CsPbBr$_3$ and Paracetamol datasets are less apparent, Adam compares well across all experiments performed. We note however that our implementation has not been heavily optimized, and further quantitative benchmarks and comparisons could be explored in future work.

\section*{Supplementary Figures}

\begin{figure}[H]
    \centering
    \includegraphics[width=0.95\linewidth]{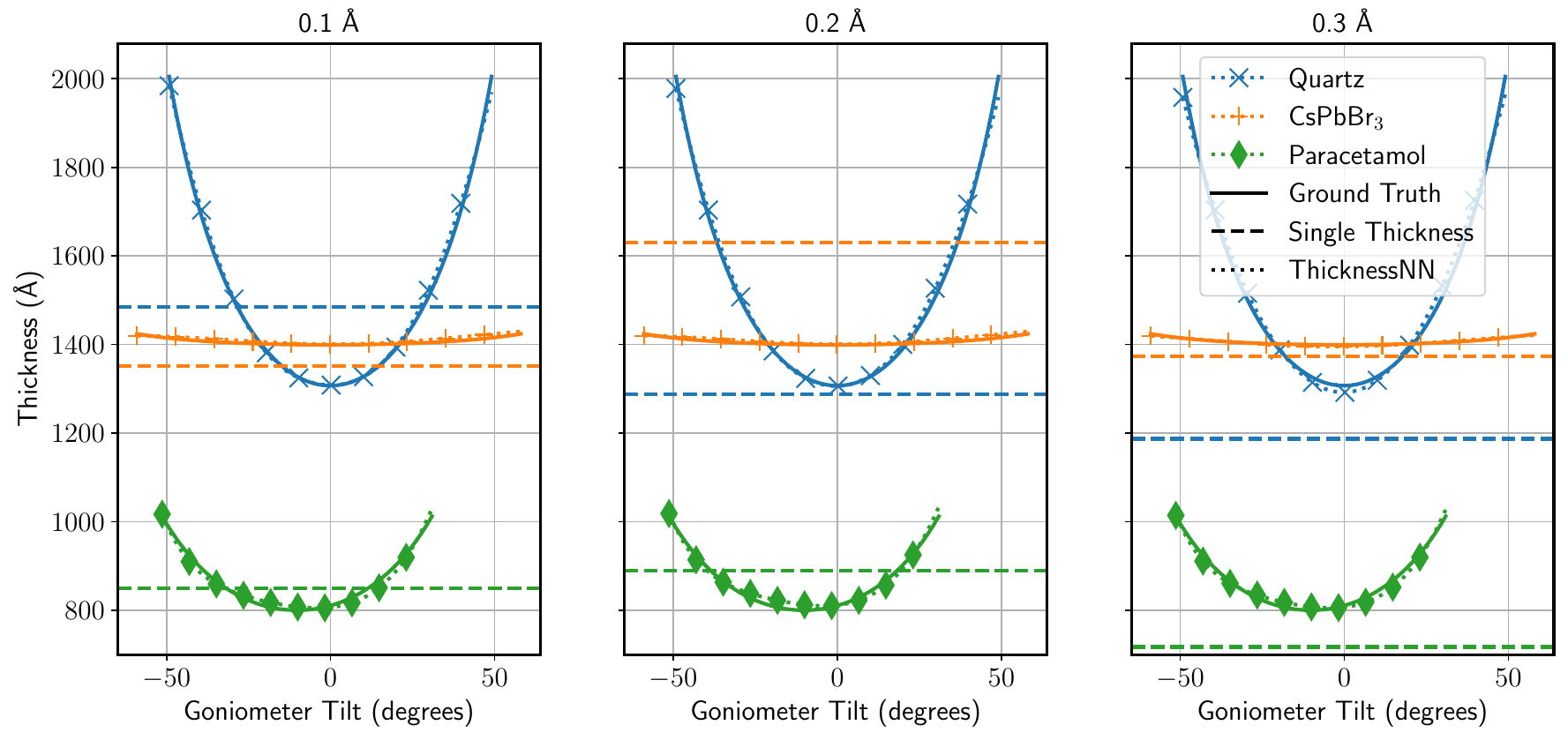}
    \caption{Ground truth synthetic thicknesses and optimized thicknesses for each displacement, comparing ThicknessNN and a single dynamical thickness. The ground truth and ThicknessNN recovered distributions are overlapping. The single dynamical thickness was found by taking the mean over the optimal thicknesses for each rotation. The optimal thickness for each rotation was found via conducting a grid search from 5 Å to 2000 Å in 2 Å increments to minimize $wR_{\text{all}}$.}
    \label{fig:thicknesses}
\end{figure}

\begin{figure}[H]
    \centering
    \includegraphics[width=1\linewidth]{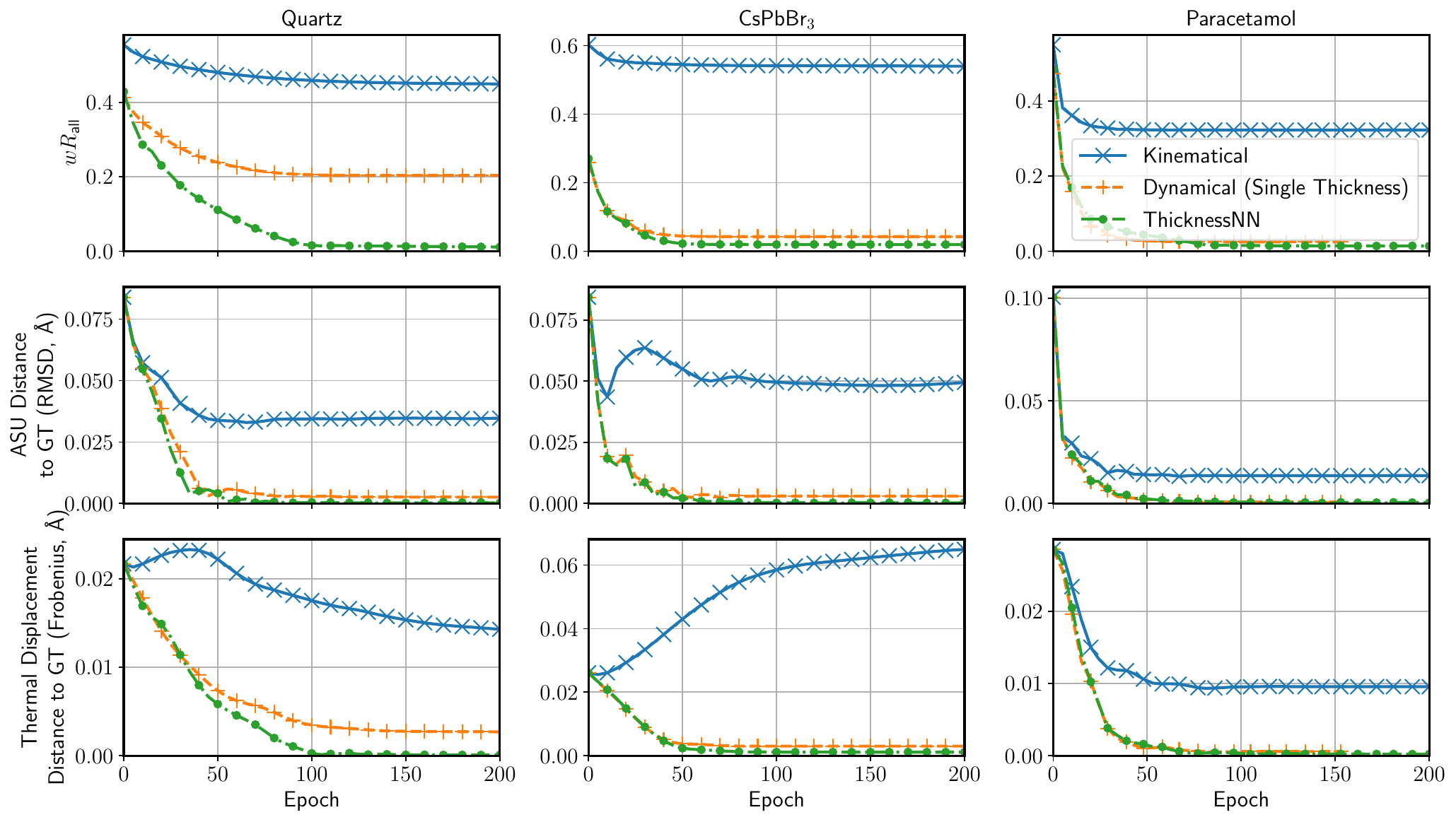}
    \caption{Refinement of synthetic crystal structures starting from 0.1 Å random maximum displacement from the ground truth. The mean diffraction intensity loss across all rotations in the dataset $wR_{\text{all}}$, the root mean squared displacement (RMSD) to the unperturbed asymmetric unit, and the Frobenium norm of difference in thermal displacement parameters are shown for each material as a function of the refinement time in epochs.}
    \label{fig:synth-refine-01}
\end{figure}

\begin{figure}
    \centering
    \includegraphics[width=1\linewidth]{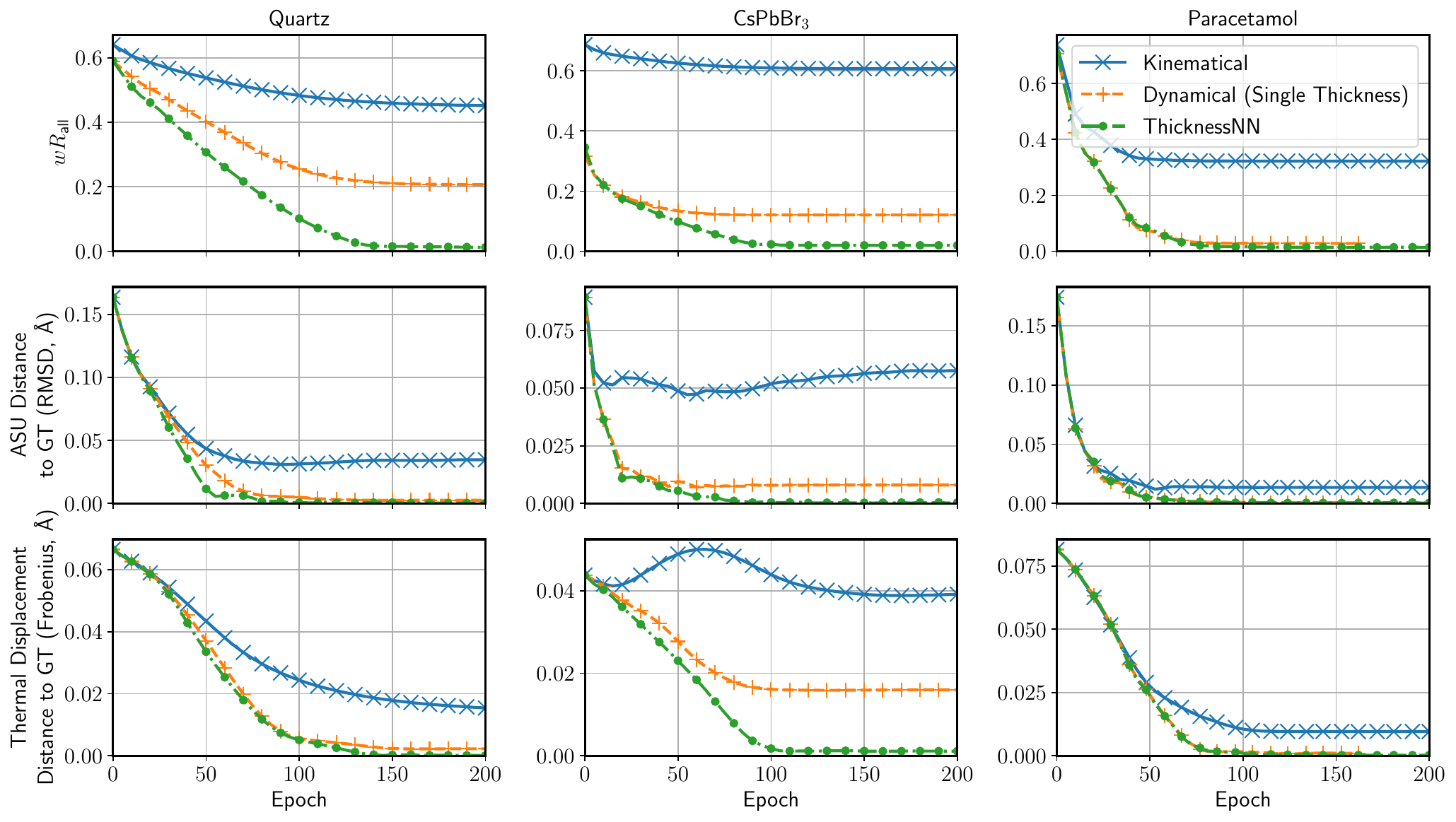}
    \caption{Refinement of synthetic crystal structures starting from 0.2 Å random maximum displacement from the ground truth. The mean diffraction intensity loss across all rotations in the dataset $wR_{\text{all}}$, the root mean squared displacement (RMSD) to the unperturbed asymmetric unit, and the Frobenium norm of difference in thermal displacement parameters are shown for each material as a function of the refinement time in epochs.}
    \label{fig:synth-refine}
\end{figure}

\begin{figure}[H]
    \centering
    \includegraphics[width=1\linewidth]{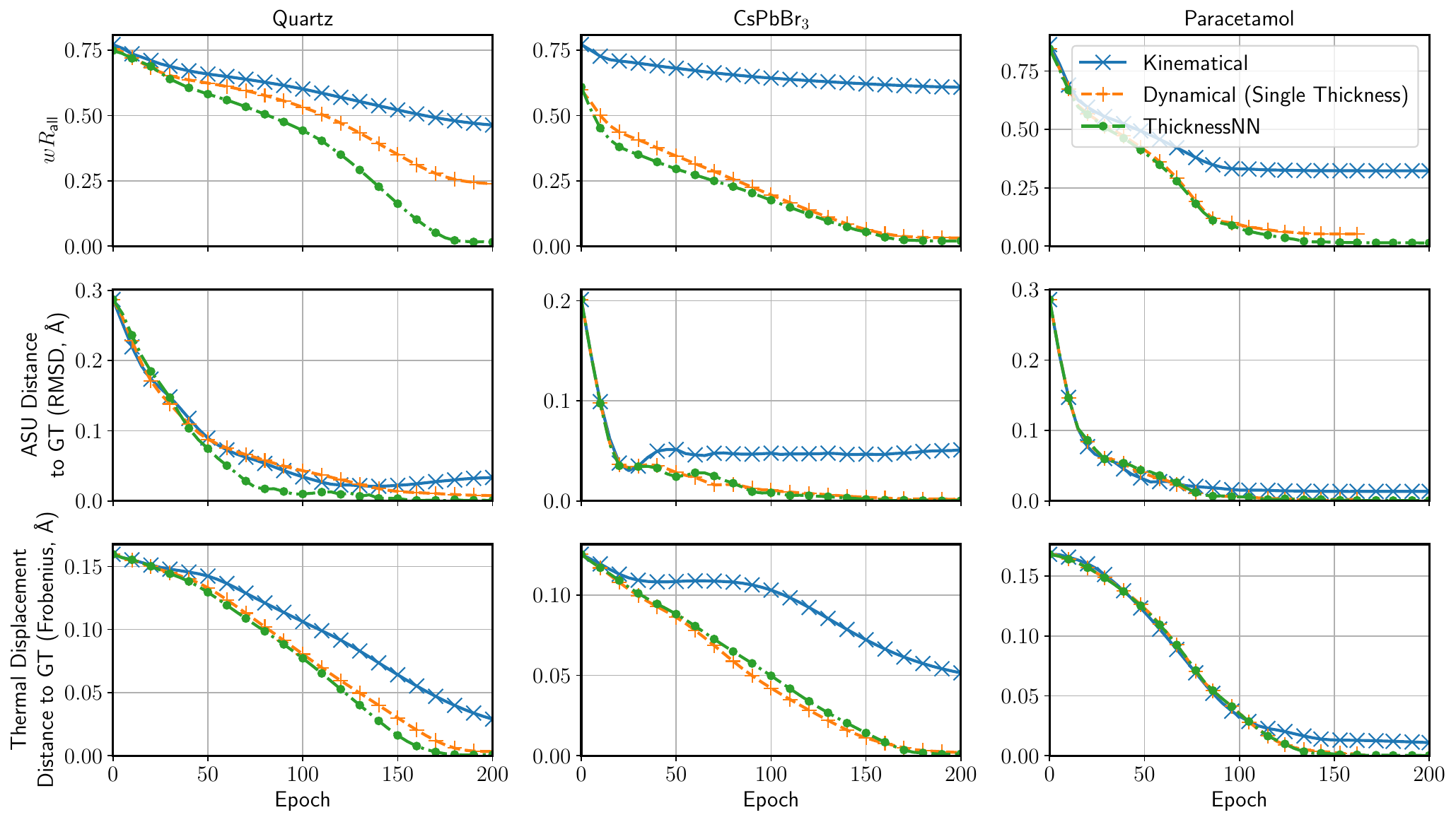}
    \caption{Refinement of synthetic crystal structures starting from 0.3 Å random maximum displacement from the ground truth. The mean diffraction intensity loss across all rotations in the dataset $wR_{\text{all}}$, the root mean squared displacement (RMSD) to the unperturbed asymmetric unit, and the Frobenium norm of difference in thermal displacement parameters are shown for each material as a function of the refinement time in epochs.}
    \label{fig:synth-refine-03}
\end{figure}

\bibliography{sn-bibliography}